\documentclass[twocolumn,showpacs,superscriptaddress,amsmath,amssymb]{revtex4}
\usepackage{graphicx}
\usepackage{dcolumn}
\usepackage{bm}

\def\bd{
\begin{document}} \def\ed{\end{document}}
\def\bmp{\begin{minipage}} \def\emp{\end{minipage}}
\def\bcc{\begin{center}} \def\ecc{\end{center}}     \def\npg{\newpage}
\def\beq{\begin{equation}} \def\eeq{\end{equation}} \def\hph{\hphantom}
\def\be{\begin{equation}} \def\ee{\end{equation}} \def\r#1{$^{[#1]}$}
\def\n{\noindent} \def\ni{\noindent} \def\pa{\parindent}
\def\hs{\hskip} \def\vs{\vskip} \def\hf{\hfill} \def\ej{\vfill\eject}
\def\cl{\centerline} \def\ob{\obeylines}  \def\ls{\leftskip}
\def\underbar#1{$\setbox0=\hbox{#1} \dp0=1.5pt \mathsurround=0pt
   \underline{\box0}$}   \def\ub{\underbar}    \def\ul{\underline}
\def\f{\left} \def\g{\right} \def\e{{\rm e}} \def\o{\over} \def\d{{\rm d}}
\def\vf{\varphi} \def\pl{\partial} \def\cov{{\rm cov}} \def\ch{{\rm ch}}
\def\la{\langle} \def\ra{\rangle} \def\EE{e$^+$e$^-$} \def\pt{p_{\rm t}}
\def\dt{\delta}   \def\sqnn{\sqrt{s_{\rm NN}}}
\def\bitz{\begin{itemize}} \def\eitz{\end{itemize}}
\def\btbl{\begin{tabular}} \def\etbl{\end{tabular}}
\def\btbb{\begin{tabbing}} \def\etbb{\end{tabbing}}
\def\beqar{\begin{eqnarray}} \def\eeqar{\end{eqnarray}}
\def\\{\hfill\break} \def\dit{\item{-}} \def\i{\item}
\def\bbb{\begin{thebibliography}{9} \itemsep=-1mm}
\def\ebb{\end{thebibliography}} \def\bb{\bibitem}
\def\bpic{\begin{picture}(260,240)} \def\epic{\end{picture}}
\def\akgt{\noindent{Acknowledgements}}
\def\fgn{\noindent{\bf\large\bf figure captions}}
\def\lan{\langle}
\def\ran{\rangle}
\def\p{\pi}
\def\ifmath#1{\relax\ifmmode #1\else $#1$\fi}%
\def\rc{\ifmath{{\mathrm{c}}}}
\def\cut{\ifmath{{\mathrm{cut}}}}
\def\rF{\ifmath{{\mathrm{F}}}}
\def\rK{\ifmath{{\mathrm{K}}}}
\def\rp{\ifmath{{\mathrm{p}}}}
\def\rt{\ifmath{{\mathrm{t}}}}
\def\LAB{\ifmath{{\mathrm{LAB}}}}
\def\cut{\ifmath{{\mathrm{cut}}}}
\def\beq{\begin{equation}}
\def\eeq{\end{equation}}
\def\us{^{(s)}}  \def\bea{\begin{eqnarray}} \def\eea{\end{eqnarray}}
\def\nbr{\nonumber} \def\e{\eta} \def\dt{\delta} \def\D{\Delta}
\def\r{\rho}
\newcommand{\cinst}[2]{$^{\mathrm{#1}}$~#2\par}
\newcommand{\crefi}[1]{$^{\mathrm{#1}}$}
\newcommand{\crefii}[2]{$^{\mathrm{#1,#2}}$}
\newcommand{\crefiii}[3]{$^{\mathrm{#1,#2,#3}}$}
\newcommand{\HRule}{\rule{0.5\linewidth}{0.5mm}}

\usepackage{color}
\newcommand{\Blue}[1]{\textcolor[named]{Blue}{#1}}
\newcommand{\blue}[1]{\textcolor[named]{Blue}{#1}}
\newcommand{\red}[1]{\textcolor[named]{Red}{#1}}
\newcommand{\violet}[1]{\textcolor[named]{Violet}{#1}}
\newcommand{\brown}[1]{\textcolor[named]{Brown}{#1}}
\newcommand{\green}[1]{\textcolor[named]{Green}{#1}}
\newcommand{\Red}[1]{\textcolor[named]{Red}{#1}}
\newcommand{\yellow}[1]{\textcolor[named]{Yellow}{#1}}
\newcommand{\magenta}[1]{\textcolor[named]{Magenta}{#1}}

\bd
\title{Trace initial interaction from final state observable \\
in relativistic heavy ion collisions}

\author{Wang Meijuan, Liu Lianshou and Wu Yuanfang}

\affiliation{Institute of Particle Physics, Huazhong Normal
University, Wuhan 430079, China}

\begin{abstract}
In order to trace the initial interaction in ultra-relativistic
heavy ion collision in all azimuthal directions, two azimuthal
multiplicity-correlation patterns
--- neighboring and fixed-to-arbitrary angular-bin correlation
patterns --- are suggested. From the simulation of Au + Au
collisions at $\sqnn=200$ GeV by using the Monte Carlo models RQMD
with hadron re-scattering and AMPT with and without string
melting, we observe that the correlation patterns change gradually
from out-of-plane preferential one to in-plane preferential one
when the centrality of collision shifts from central to
peripheral, meanwhile the anisotropic collective flow $v_2$ keeps
positive in all cases. This regularity is found to be model and
collision energy independent. The physics behind the two opposite
trends of correlation patterns, in particular, the presence of
out-of-plane correlation patterns at RHIC energy, are discussed.
\end{abstract}


\maketitle

The data from current relativistic heavy ion experiments show that
a new form of matter --- quark-gluon plasma (QGP) has been
produced at RHIC~\cite{qgp, gulassys}. The anisotropic collective
flow $v_2$ is supposed to provide us the information at the early
stage of collision. The successful hydrodynamic
description~\cite{mv2} on observed mass dependence of $v_2$ at low
transverse-momentum range ($p_{\rm T} < 2$GeV ) shows that the
observed dense matter behaves like a perfect fluid rather than an
ideal gas, and is, therefore, referred to as sQGP, although
hydrodynamics can still not well fit all the observed data in the
range~\cite{break}

A complete information about initial interaction and evolution is
very important for correctly understanding the properties of the
formed matter. The anisotropic collective flow~\cite{aflow} $v_2$
is the second Fourier coefficient of the transverse-momentum
distribution of final state particles. It only globally
characterizes the direction and strength of anisotropic
distribution. The intrinsic interaction (correlation) of final
state particles is absent in the measure.

Another related measure at the market is the 2-particle azimuthal
correlation~\cite{phenix,2par}. It concerns the average
correlation of two particles separated by a certain angle, no
matter where the two particles are in the whole azimuthal space.
The 3-particle, or 4-particle, azimuthal correlations focus on the
same kind of correlations. How the particles in different
azimuthal directions interact with each other can not be drawn
from them either.

Newly suggested two spatial-dependent correlation patterns,
neighboring and fixed-to-arbitrary bin correlation
patterns~\cite{wu-pre}, provide information on how the particles
in different cells of phase space are correlated. They measure in
the whole phase space the two typical correlations, i.e.,
correlations between local particles and between particles with a
certain distance in phase space. Therefore, they are good
observable for tracing the initial interaction among different
azimuthal directions in relativistic heavy ion collision.

In this letter, we will first introduce these two correlation
patterns into relativistic heavy ion collision. Then, using the
Monte Carlo models RQMD~\cite{rqmd} and AMPT~\cite{ampt}, we
demonstrate that the out-of-plane correlation pattern coexists
with the in-plane one at RHIC energy, although the anisotropic
elliptic flow $v_2$ keeps positive there. Finally, the physical
origin which generates the two opposite correlations
--- out-of-plane and in-plane preferential correlation patterns is discussed.

It is well-known that the general 2-bin correlation is defined as
\beq C_{m_1,m_2}= \frac{\langle n_{m_1} n_{m_2} \rangle}{\langle
n_{m_1} \rangle \langle n_{m_2} \rangle}-1, \eeq \noindent where
$m_1$ and $m_2$ are the positions of the two bins in phase space and
$n_m$ is the measured content in the $m$th bin. If there is no
correlation between particles in the observed window, $C_{m_1,m_2}$
vanishes.

We divide the $2\pi$ azimuthal angle equally into $M$ bins and
specify $n_m$ as the multiplicity in the $m$th angular bin. If we
let $m_1=m$ and $m_2=m+1$, $C_{m_1,m_2}$ is reduced to the {\it
neighboring angular-bin correlation pattern}, \beq
C_{m,m+1}=\frac{\langle n_{m}n_{m+1}\rangle}{\langle
n_{m}\rangle\langle n_{m+1}\rangle}-1. \eeq \noindent If we fix
the position of one bin, $m_1=m_0$, and vary the left one by $m$,
it becomes the {\it fixed-to-arbitrary angular-bin correlation
pattern}, \beq C_{m_0,m}=\frac{\langle
n_{m_0}n_{m}\rangle}{\langle n_{m_0}\rangle\langle
n_{m}\rangle}-1. \eeq It is clear that these two correlation
patterns measure {\it at various azimuthal positions} how the
local particles and the particles separated by a certain angle
correlate with each other.

In order to apply these two measures to current relativistic heavy
collisions, we choose the RQMD and AMPT models as examples. As we
know, the RQMD (relativistic quantum molecular dynamics) with
re-scattering is a hadron-based transport model~\cite{rqmd}. The
final hadron interactions are implemented in the model by hadron
re-scattering. Although the anisotropic collective flow produced
by the model is much smaller than the observed data at RHIC, we
can still see how the suggested correlation patterns behave for
this kind of transport model.

In contrary to the RQMD model, the AMPT is a "multi-phase"
transport model, with hadron and parton interactions both included
in. In the default AMPT, the transport in parton level is only
partly, and is accompanied by string evolution. While in the AMPT
with string melting, the parton level transport is fully taken
into account, and reproduces the observed anisotropic collective
flow at RHIC~\cite{ampt}. So, using this model we will see how the
azimuthal interaction changes when the anisotropic expansion
develops from partly to fully.

We generate 249,824, 204,004, and 686,278 events from RQMD and
AMPT with and without string melting, respectively. The centrality
dependence of neighboring angular-bin multiplicity correlation
patterns for Au + Au collisions at $\sqnn=200$ GeV are shown in
Fig.~1, where the results from the RQMD with re-scattering, from
the AMPT with and without string melting are presented in the
first to third columns, respectively. As an example, the
fixed-to-arbitrary angular-bin multiplicity correlation patterns
for Au+Au collisions at the same energy from RQMD with
re-scattering are shown in Fig.~2. Here we partition the whole
azimuthal range $2\pi$ uniformly into 50 equal size angular bins
and in the first to third columns of Fig.~2 the positions of the
fixed bins are located at $m_0=1, 6, 12$, corresponding to the
azimuthal angles $\phi\cong0, \frac{\pi}{4}, \frac{\pi}{2}$,
respectively. The errors are statistical only and most of them are
smaller than the symbol size in these two and following figures.
In this analysis, the $\phi =0$ refers to the direction of
reaction plane in nuclear collisions. In real experimental data
analysis, it has to be determined event-by-event.

\begin{figure}
\includegraphics[width=3.4in]{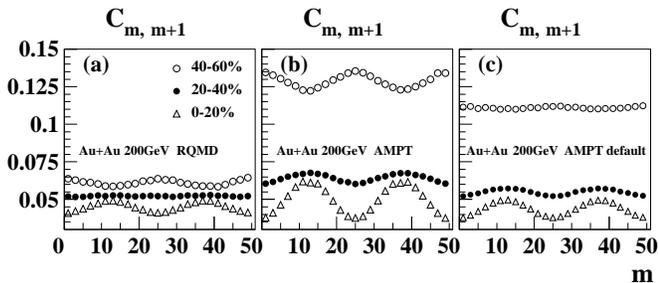}
\caption{\label{Fig. 1} The centrality dependence of neighboring
angular-bin correlation patterns for Au + Au collisions at 200 GeV
from the RQMD with re-scattering (first column), AMPT with (second
column) and without (third column) string melting.}
\end{figure}

\begin{figure}
\includegraphics[width=3.4in]{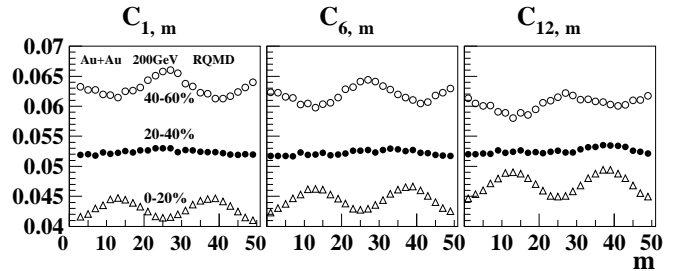}
\caption{\label{Fig. 2} The centrality dependence of
fixed-to-arbitrary angular-bin correlation patterns for Au + Au
collisions at 200 GeV from the RQMD with re-scattering.}
\end{figure}

We can see from Fig.~1 that the RQMD with re-scattering, the AMPT
with and without string melting give qualitatively the same
centrality dependence of azimuthal correlation patterns. The
correlation patterns are $\cos^2\phi$ like in peripheral
collisions, then turn to flat in mid-central collisions, and
become $\sin^2\phi$ like in near-central collisions. Here, we
present only three centrality ranges to show their typical
behavior. In fact, the correlation pattern changes gradually from
$\cos^2\phi$ like to $\sin^2\phi$ like with centrality. In the
default AMPT the $\sin^2\phi$ like pattern is very weak at the
centrality 40-60\%, and becomes more obvious in more peripheral
collision with centrality $60-80\%$ (not shown here). It is clear
that two opposite trends dominate in peripheral and near-central
collisions, respectively. In the mid-central collisions, the two
trends turn to balance and the correlations become equal in all
directions. Moreover, these important characteristics are
independent of the specific assumptions implemented in the three
models, in particular independent of the hadronization schemes
assumed in these models.

It is interesting and worthwhile to study these two opposite
trends in detail. In peripheral collisions, corresponding to the
centrality at $40\sim 60\%$, both correlation patterns show
$\cos^2\phi$ - like patterns. The strengthes of the correlations
are the largest at $\phi=0, \pi, 2\pi$ in neighboring angular-bin
correlation pattern and are varying in fixed-to-arbitrary angular
correlation pattern
--- the nearer the fixed angular bin to $\phi=0$ or $\pi$, the
larger the correlation strength. It tells us that the final state
particles in the directions of in-plane ($x$-$z$ plane, with
$z$-axis corresponding to the beam direction and $x$ axis along
the impact parameter as shown in Fig.~3(a)) have the strongest
neighboring correlations, and have also the strongest correlation
with the particles in other directions. But the strength of the
correlation decreases when the fixed angle goes away from the
direction of in-plane as shown in the first to third columns of
Fig.~2. These results show that the behavior of particles in
in-plane directions has the strongest influence to the behavior of
particles in all other azimuthal directions, in contrast with the
fixed-to-arbitrary angular-bin correlation patterns in
hadron-hadron collisions, where the back-to-back correlations are
always the strongest one~\cite{wang-hangzhou}. These characters
are similar to that of in-plane flow resulted from anisotropic
expansion, which makes expansion privileged, or correlation
stronger, in the directions of in-plane.

On the contrary, approaching to the near-central collision,
corresponding to the centrality at $0\sim 20\%$, the two
correlation patterns show $\sin^2\phi$-like patterns. The
strongest correlations are located in both ${\pi\over 2}$ and
${3\pi\over 2}$, i.e. out-of-plane ($y$-$z$ plane) directions,
instead of $0$ and $\pi$ (in-plane directions), and the particles
in the out-of-plane directions have also the strongest correlation
with the particles in other directions. It indicates that the
behavior of particles in out-of-plane directions has the strongest
influence to the behavior of particles in all other azimuthal
directions. These characters are opposite to the in-plane
flow~\cite{aflow, starflow}, but are similar to the out-of-plane
one, which has been observed at Bevalac and SIS
energies~\cite{squeeze-e}. Such an out-of-plane flow was explained
as that, before the anisotropic expansion, the participant
nucleons, which are compressed in the overlap zone, cannot escape
in the reaction plane ($x$-$z$) direction due to the presence of
the spectator nucleons (squeeze-out effect~\cite{squeeze,
reviews}), producing out-of-plane flow. However, when Lorentz
contraction effect becomes significant at ultra-relativistic
collisions, the spectators will leave very soon from participant
zone~\cite{nu}. The initial squeeze-out effect is supposed to be
hardly realizable at RHIC energies~\cite{reviews}. Therefore, the
observed out-of-plane correlation pattern is not due to the
initial squeeze-out effect.

It is clear that the anisotropic expansion and the late
hadronization are impossible to produce stronge correlations in
out-of-plane directions. Only the initial source anisotropy is
preferential in these directions. The anisotropy of the initial
overlap interaction area at non-central collisions results in a
larger initial number of participant nucleons in the out-of-plane
directions, which in turn generates stronger interaction in these
directions.

\begin{figure}
\includegraphics[width=3.2in]{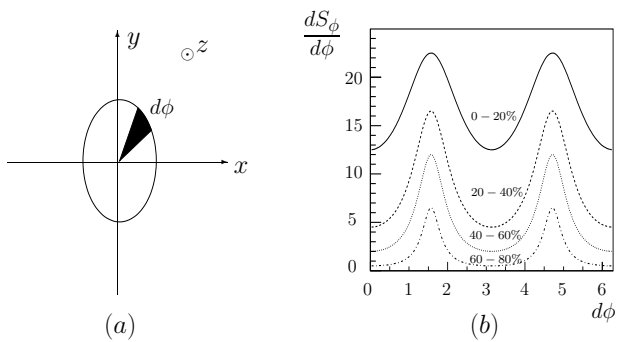}
\caption{\label{Fig. 3}(a) An angular bin in the overlap zone,
which is approximated as an ellipse in coordinate space; (b) The
azimuthal distribution of overlap area at four centralities. }
\end{figure}

Let us estimate the azimuthal distribution of participants in the
overlap zone and see whether it has the same centrality dependence
as the out-of-plane correlation patterns. If we neglect the
difference of density in radial direction and approximate the
overlap zone as an ellipse, the number of participants should be
approximately proportional to the geometrical area of the sector
in a given angular bin $\d\phi$ as shown in Fig.~3(a), i.e.,
\beq\frac{\d N_{\rm part}}{\d\phi}\propto\frac{\d
S_{\phi}}{\d\phi}=\frac{1}{2}\rho^2=
\frac{a^2b^2}{2(b^2\sin^2\phi+a^2\cos^2\phi)}, \eeq\noindent where
$N_{\rm part}$ is the number of participant nucleons, $S_{\phi}$
the overlap area in the angular bin $d\phi$, $\rho$ the radius of
the overlap zone in the considered direction. $a$ and $b$ are the
semi-major and semi-minor axes of the overlap ellipse and can be
deduced from the impact parameter $b_{\rm im}$ and the radius $R$
of colliding nucleus as $a=\sqrt{R^2-b_{\rm im}^2/4}$, $b=R-b_{\rm
im}/2$.

In Fig.~3(b), the azimuthal distribution of participants at four
centralities are presented. We can see from the figure that the
azimuthal distribution of participants indeed behaves similar to
$\sin^2\phi$, mimic the neighboring angular bin correlation
pattern for near-central collisions shown in Fig.~1. The largest
number of participant nucleons are located at $\frac{\pi}{2}$ and
$\frac{3\pi}{2}$. The amplitude of the distribution is small in
peripheral collisions, increases with the increasing of
centrality, and reaches its maximum amplitude at the centrality
$20-40\%$. It goes down slowly with the further increasing of
centrality. The amplitude of the distribution will vanish in head
on collisions as there is no difference in all directions of
overlap area in such collisions. We indeed observed that the
correlation patterns tend to flat at very central (centrality at
$0-5\%$) collisions in the AMPT with string melting. Such
centrality dependence of the azimuthal distribution of
participants can provide us a good understanding in the centrality
dependency of the out-of-plane correlation patterns.

The observed correlation patterns present the interactions between
the particles in different azimuthal directions. They are resulted
from the competition of all kinds of interactions after collision,
in particular, the initial interaction and subsequent anisotropic
expansion. As long as the system is not fully thermalized, the
initial out-of-plane preferential interaction will compete with
the subsequent anisotropic expansion. In peripheral collisions,
the overlap zone is small and so is the number of participant
nucleons, but the difference between the minor and major axes of
overlap ellipse is large, and so is the difference of pressure
gradient. In this case the anisotropic expansion dominates the
final observable, and the effects of initial interaction in
correlation patterns are hidden. In near-central collisions, the
overlap zone becomes large and the difference between minor and
major axes of ellipse is small, so that the initial interactions
are strong enough to show themselves up in final observable. This
is why the out-of-plane correlation patterns appear at
near-central collisions.

Therefore, the azimuthal multiplicity-correlation patterns provide
us the information {\it not only } on the subsequent anisotropic
expansion {\it but also} on the full initial interactions, which
are sensitive to the anisotropic shape of initial overlap zone.

We can further see from Fig.~1 that (1) the values of out-of-plane
correlation patterns are almost the same in different models; (2)
the values of in-plane ones vary greatly from one model to
another; (3) the oscillation amplitudes of both the two opposite
correlation patterns are the biggest in the AMPT with string
melting. The reason of (1) is that the anisotropy of initial
interaction is mainly determined by the geometry of initial
interaction zone, which is the same in three models. Meanwhile,
the subsequent anisotropic expansion is realized by the late
transportation, which is very different in three models, and so
are the results of (2). In addition, at the same collision energy,
the strength of initial interaction also depends on the
interaction mechanism~\cite{rqmd,ampt}. The anisotropic expansion
is built up on the initial out-of-plane preferential interaction.
A stronge initial interaction with full transporting results in
large correlations in both in-plane and out-of-plane directions,
i.e., (3).

In order to see the energy dependence of the correlation pattern,
the neighboring angular-bin correlation patterns for Au + Au
collisions produced by RQMD with hadron re-scattering at 60 GeV
and 5 GeV are shown in Fig.~4(a) and (b) respectively. It is clear
that the higher the collision energy, the bigger the values and
amplitudes of out-of-plane and in-plane correlation patterns. We
can also see from the figure that the presence of out-of-plane and
in-plane correlation patterns is independent of collision energy,
but dependent on the centrality only.

However, if the initial eccentricity in coordinate space has not
been transformed into, or inherited to, the transverse-momentum of
final state particles, two opposite correlation patterns will be
absent in the final observable. This is the case for RQMD without
hadron re-scattering, where the transverse-momentum distribution
is isotropic and correlation patterns are flat for all
centralities.

\begin{figure}
\includegraphics[width=3.4in]{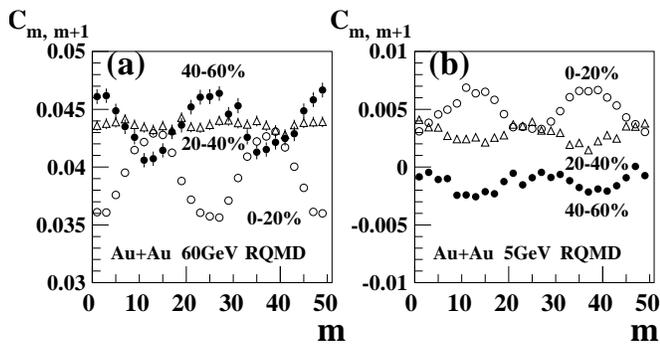}
\caption{\label{Fig. 4} The centrality dependence of neighboring
angular-bin correlation pattern for Au + Au collision at (a) 60
GeV and (b) 5 GeV from the RQMD with re-scattering.}
\end{figure}
As we know, anisotropic collective flow, $v_2$, keeps positive in
RQMD with re-scattering and in the AMPT with and without string
melting at 200GeV, 60 GeV and 5 GeV, which are in consistent with
the corresponding data. The results from the suggested patterns
show that even if we observe in-plane preferential
transverse-momentum distribution, it does not mean that the
correlations, or interactions, have the same preferential
directions.

To the summary, we suggest to apply the two azimuthal
multiplicity-correlation patterns in ultra-relativistic heavy ion
collisions. They are neighboring and fixed-to-arbitrary
angular-bin multiplicity-correlation patterns. From the simulation
of Au + Au collisions at 200, 60 and 5 GeV by using the MC models
RQMD with hadron re-scattering and AMPT with and without string
melting, we observe, model and energy independently, that the
correlation patterns change gradually from out-of-plane
preferential one to in-plane preferential one when the centrality
of collision decreases from central to peripheral, while $v_2$
keeps positive in all cases. The in-plane preferential correlation
patterns are resulted by anisotropic expansion, and the
out-of-plane ones are found to be likely caused by the initial
interaction due to the eccentricity of initial overlap zone. The
experimental observation of the out-of-plane correlation patterns
predicted in the present letter will offer us a better
understanding for the initial pre-thermalization interaction and
its competition with the subsequent collective
expansion~\cite{break, eccen}and is, therefore, called for.

Wu Yuanfang would thank Dr. Nu Xu for his stimulating comments. We
are grateful for the financial supports from the NSFC of China
under projects: No. 90503001, 10610285, 10775056.

\ed